\documentclass[prl,amssymb,twocolumn,showpacs,supersriptaddress]{revtex4}
\usepackage{amssymb}
\usepackage{graphicx}
\usepackage{bm}

\begin{document}

\title{Tailoring population inversion in 
Landau-Zener-St\"uckelberg interferometry
of flux qubits}

\author{Alejandro Ferr\'on}
\affiliation{Instituto de Modelado e Innovaci\'on Tecnol\'ogica (CONICET-UNNE), 3400 Corrientes}
\affiliation{Centro At{\'{o}}mico Bariloche and Instituto Balseiro, 8400 San Carlos de Bariloche, R\'{\i}o Negro, Argentina.}
\author{Daniel Dom\'{\i}nguez}
\author{Mar\'{\i}a Jos\'{e} S\'{a}nchez}
\affiliation{Centro At{\'{o}}mico Bariloche and Instituto Balseiro, 8400 San Carlos de Bariloche, R\'{\i}o Negro, Argentina.}

\begin{abstract}
We distinguish different mechanisms for population inversion in flux qubits 
driven by dc+ac magnetic fields. We show that for driving amplitudes
such that there are Landau-Zener-St\"uckelberg
interferences,  it is possible to have population inversion 
solely mediated by the environmental bath.
Furthermore, we find 
that the degree of population inversion
can be controlled by tailoring 
a resonant frequency $\Omega_p$ in the
environmental bath. 
To observe these effects experiments should be performed
for long driving times after full relaxation. 
\end{abstract}


\maketitle

Population inversion, where the most highly populated state is
an excited state, is among the most interesting phenomena  in
maser and laser physics \cite{lasers}. 
The usual mechanism  to get  population inversion (PI) requires 
driven  quantum systems with three or more energy levels
\cite{lasers,lukens,nori2007,astafiev}. 
Interesting systems to study PI are
`artificial atoms' made with 
mesoscopic Josephson devices
\cite{revqubits,nori2011}.
Among them, 
a well known  circuit is the flux qubit (FQ) \cite{qbit_mooij,chiorescu}. 
When driven by  a dc+ac magnetic flux,
it has transitions between energy levels at  avoided crossings.
A rich structure of Landau-Zener-St\"uckelberg (LZS)
interferences  \cite{shevchenko} combined with multi-photon resonances 
\cite{oliver,valenzuela,izmalkov} is observed.
LZS interference patterns have also been measured in
charge qubits \cite{cqubits}, Rydberg atoms \cite{yoakum}, ultracold molecular gases \cite{mark},
optical lattices \cite{kling} and single electron spins \cite{huang}.
In FQ, diamond-like  patterns were found for large ac amplitudes,
from which the energy level spectrum has been reconstructed \cite{valenzuela}.
Population inversion was observed in the `second diamond',  {\it i. e.}
for ac amplitudes that  excite to the third and four energy levels
through Landau-Zener transitions. PI was  also measured 
in another flux qubit device  driven at large frequencies \cite{sun}.
Here we will show that an even more notable effect is awaiting to be observed in these
systems if the ac driving pulses are applied for  longer time scales: 
 PI could be observable  even 
for ac amplitudes where only the two lowest levels of the FQ participate.
In the following we will explain, through realistic time-dependent numerical simulations,
(i) how this type of PI can occur, (ii) why it has not
been observed experimentally yet and (iii) how the occurrence and the degree 
of PI can be tailored by changing the structure of the environmental bath.

The FQ  consists of a   superconducting ring with three Josephson
junctions \cite{qbit_mooij} enclosing a magnetic flux $\Phi=
f\Phi_0$ ($\Phi_0=h/2e$) with 
phase differences
$\varphi_1$, $\varphi_2$ and
$\varphi_3=-\varphi _1 +\varphi _2 -2\pi f$.
Two of the junctions have  coupling energy,
$E_J$, and capacitance, $C$,
and the other
has  $E_{J,3}=\alpha E_J$ and $C_3=\alpha C$. 
The hamiltonian is: \cite{qbit_mooij}
\begin{equation}
\label{ham-sys}
{\cal H}=-2E_C\left(\frac{\partial^2}{\partial\varphi_t^2}+
\frac{1}{1+2\alpha}\frac{\partial^2}{\partial\varphi_l^2}\right)
+E_JV(\varphi_t,\varphi_l)\; ,
\end{equation}
with
$\varphi_t=(\varphi_1+\varphi_2)/2$, 
$\varphi_l=(\varphi_1-\varphi_2)/2$,
$E_C=e^2/2C$ and $V(\varphi_t,\varphi_l)= 2+\alpha -
2\cos\varphi_t\cos \varphi_l - \alpha \cos (2\pi f+2\varphi _l)$.
The FQ has several discrete levels with eigenenergies $E_i$ 
and eigenstates 
$|\Psi_i\rangle$
which depend on $f$, 
$\alpha$ and  $\eta=\sqrt{8E_C/E_J}$.
Typical experiments  have $\alpha \sim 0.6-0.9$ and $\eta\sim 0.1-0.6$ 
\cite{chiorescu,oliver,valenzuela}.
For $\alpha \ge 1/2$ and
$|f- 1/2| \ll 1$, the potential $V(\varphi_t,\varphi_l)$
has the shape of a double-well with two minima along the $\varphi_l$ direction. 
In this regime the system can be
operated as a
quantum bit \cite{qbit_mooij,chiorescu} and  approximated
by a two-level system (TLS) \cite{qbit_mooij,ferron}. 
When  driven by a  magnetic flux 
$f(t)=f_{dc}+f_{ac}\sin(\omega t)$, the hamiltonian is 
time periodic ${\cal H}(t) = {\cal H}(t + \tau)$, with $\tau=2\pi/\omega$.
In  the Floquet formalism, which allows to treat  periodic forces
of arbitrary strength and frequency \cite{floquet,fds},
the solutions of the    Schr\"odinger equation are of the
form  $|\Psi_\beta(t)\rangle=e^{i\varepsilon_\beta t/\hbar}|\Phi_\beta(t)\rangle$, where
the  Floquet states $|\Phi_\beta(t)\rangle$
satisfy $|\Phi_\beta(t)\rangle$=$|\Phi_\beta(t+ \tau)\rangle = 
\sum_k |\Phi_\beta^{k} \rangle e^{-ik\omega t}$, and
are eigenstates of the equation
$[{\cal H} (t)- i \hbar \partial/\partial t ] |\Phi_\beta(t)\rangle= \varepsilon_\beta |\Phi_\beta(t)\rangle$,
with $\varepsilon_\beta$ the associated quasi-energy.

Experimentally, the system is 
affected by the electromagnetic
environment that  introduces 
decoherence and relaxation  processes.
A  standard theoretical
model is to linearly couple
the system to a bath
of harmonic oscillators \cite{hanggi,milenas,grifonih,caspar}
with a spectral density
$J(\omega)$ and   equilibrated at  temperature $T$.
For the FQ, the dominating source of decoherence is flux noise,
in which case the bath degrees of freedom couple with
the system variable $\varphi_l$ (see \cite{caspar}).
For  weak coupling  (Born approximation) 
and fast bath relaxation (Markov approximation),
a Floquet-Born-Markov master  equation  
for   the reduced density matrix $\hat\rho$
in the Floquet basis,
$\rho_{\alpha\beta}(t)=\langle\Phi_\alpha(t)|\hat\rho(t)|\Phi_\beta(t)\rangle$,
can be obtained \cite{hanggi}:
\begin{equation}
\label{drho}
\frac{d\rho_{\alpha\beta}(t)}{dt}=-\frac{i}{\hbar}
(\varepsilon_\alpha-\varepsilon_\beta)\rho_{\alpha\beta}(t) +
\sum_{\alpha'\beta'}{\it L}_{\alpha'\beta'\alpha\beta}(t) \; \rho
_{\alpha'\beta'}(t) \; .
\label{rw1}
\end{equation}
The  ${\it L}_{\alpha'\beta'\alpha\beta}(t)$ are approximated
by their average over  one period $\tau$ \cite{hanggi}
since the time scale $t_r$ for full relaxation satisfies $t_r\gg \tau$, 
obtaining 
${\it L}_{\alpha'\beta'\alpha\beta}= R_{\alpha'\beta'\alpha\beta}
+R_{\beta'\alpha'\beta\alpha}^*-\sum_\eta \left( \delta_{\beta \beta'}
R_{\alpha'\alpha\eta\eta}+ \delta_{\alpha \alpha'}R_{\beta'\beta\eta\eta}^* \right)$.
The rates 
$R_{\alpha'\beta'\alpha\beta} = \sum_{n } g_{\alpha \alpha'}^n\Gamma_{\alpha \alpha'}^n\Gamma_{\beta' \beta}^{-n}$
are sums of $n$-photon exchange terms,
with $g_{\alpha \alpha'}^n=J(\frac{\varepsilon_{\alpha\alpha'}^n}{\hbar}) \left(\coth
(\frac{\varepsilon_{\alpha\alpha'}^n}{2 k_{B} T})- 1\right)$,
$\varepsilon_{\alpha\alpha'}^n=\varepsilon_\alpha-\varepsilon_{\alpha'}+n\hbar \omega$,
and $\Gamma_{\alpha\alpha'}^n=\sum_{k}\langle\Phi_\alpha^{k}| \varphi_l |\Phi_{\alpha'}^{k+n}\rangle$.
This formalism  has been extensively
employed to study relaxation and decoherence for
time dependent periodic evolutions in double-well potentials
and in  TLS \cite{hanggi,milenas,grifonih}.
Here we  use it to model the ac driven FQ
for the full hamiltonian of Eq.~(\ref{ham-sys}),
beyond the TLS approximation.
We calculate
the Floquet states 
$|\Phi_\beta^{k} \rangle$ and quasienergies $\varepsilon_\beta$ \cite{fds}, and then
we integrate numerically Eq.~(\ref{drho}), obtaining $\rho_{\alpha\beta}(t)$ as 
a function of $t$.

\begin{figure}[h]
\begin{center}
\includegraphics[width=0.9\linewidth,clip]{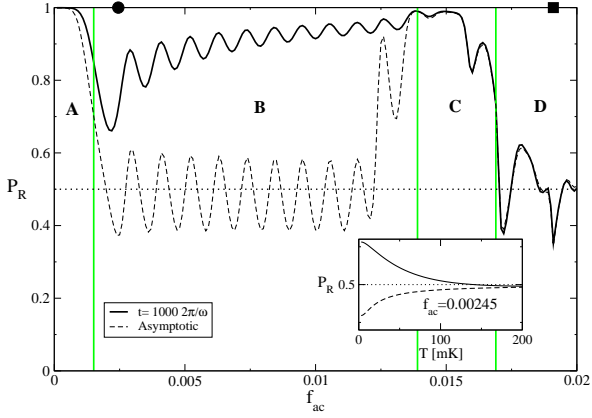}
\end{center}
\caption{
Population $P_R$ as a function of the driving amplitude $f_{ac}$. 
Continuous line: $P_R(t=1000 \tau \sim t_{\rm exp})$. 
Dashed line:  asymptotic  average population $\overline P_R$.
Inset:  $P_R$ vs. $T$ for a  driving amplitude $f_{ac}=0.00245$.
The calculations were performed for $f_{dc}=0.50151$, 
$\omega= 0.003 E_J/\hbar$ and
an ohmic bath  with $\gamma=0.001$ at $T=20 \rm{mK}$ (for $E_J/h\approx300{\rm GHz}$).
Vertical lines separate regimes A, B, C, D described in the text.
Black circle: $f_{ac}$  corresponding to the inset, Fig.2(a) and Fig.3.
Black square: $f_{ac}$ corresponding to Fig.2(b).
} \label{f1}
\end{figure}

We start by considering a FQ coupled to a bath
with an ohmic spectral density $J(\omega)=\gamma\omega$.
We take $\gamma=0.001$, corresponding to weak dissipation as in \cite{valenzuela},
and the bath  at $T= 0.0014 E_J/k_B$ ($\sim 20{\rm mK}$
for $E_J/h \sim 300{\rm GHz}$). 
The FQ device parameters are $\alpha=0.8$ and $\eta=0.25$.
The static field is taken as $f_{dc}  = 0.50151$, 
and for the driving  microwave field we choose 
 $\omega=0.003E_J/\hbar\sim 900\rm{MHz}$ 
and different amplitudes $f_{ac}$.
The initial condition is the ground state $|\Psi_0\rangle$ of
the static hamiltonian $H_0\equiv{\cal H}(f_{dc})$.

Experimentally, the probability of having a state of positive or negative
persistent current in the FQ is measured \cite{chiorescu}. The probability
of a positive current measurement (``right'' side of the double-well potential) can
be calculated as
$P_R(t)={\rm Tr}(\hat\Pi_R \hat \rho(t))$, with $\hat\Pi_R$ the operator that projects
wave functions on the $\varphi_l>0$ subspace \cite{fds}.
For a static field  $f_{dc}\gtrsim 1/2$, the
ground state has $P_R\approx 1$.
In Fig.1 we show, as a function of $f_{ac}$, $P_R(t)$
after  a driving time $t=1000\tau$,  which
is similar to the time scale  $t_{\rm exp}$ of the experiments  \cite{valenzuela}.
We also plot the asymptotic ${\overline P_R}\equiv 
{\rm lim}_{t\rightarrow\infty}\langle P_R(t)\rangle_\tau$, averaged
over one period $\tau$.
We find different regimes:
(A) For  small $f_{ac}$,  $P_R \sim 1$, since the system is slightly perturbed
from the ground state.
(B) When further increasing $f_{ac}$, new regimes are found whenever
the extreme driving amplitudes $f_{dc}\pm f_{ac}$ reach an avoided
crossing in the energy level spectrum $\{E_i(f))\}$.
Since the slopes $\frac{-dE_i}{df}$ are proportional to the 
average loop current of the FQ, at avoided crossings
there is  interference between  ``positive'' and ``negative'' loop current
states,  
which results in LZS oscillations \cite{shevchenko}
in the dependence of $P_R$ with $f_{ac}$.
The first case is found  when  the avoided
crossing  at $f=1/2$ between
the ground state level $E_0$ and the first excited level $E_1$ 
 is reached, and the  transfer of population to the $E_1$ level lowers $P_R$.
The FQ of \cite{oliver,valenzuela} has  a  decoherence time $t_\phi\sim 20{\rm ns}>\tau$  
and a large `interwell` relaxation time $t_r\sim 100\mu{\rm s}$.
Due to this time scale separation 
($t_\phi < t_{\rm exp} \ll t_r$), a TLS model with classical
noise, valid for $t<t_r$, can explain the experimentally observed LZS oscillations
of $P_R$ vs. $f_{ac}$, where the minima correlate with the zeros of
Bessel functions $J_n(af_{ac}/\omega)$ 
($a$ a normalization constant) \cite{oliver,shevchenko}.
The results of Fig.1 for $t=1000 \tau\sim t_{\rm exp}$ are in agreement with this finding.
However,  the finite time $P_R(t_{\rm exp})$ is very different from the asymptotic 
${\overline P_R}$, that even shows population inversion, ${\overline P_R} < 1/2$.
(C) At higher values of $f_{ac}$ the avoided level crossing between 
$E_1$ and $E_2$ is reached and the
ground level $E_0$ is repopulated  due to  fast
$E_2\rightarrow E_0$ `intrawell' transitions (the corresponding states  have the
same sign of the average loop current).
Thus, $P_R$ increases close to $1$ (`cooling' effect, see \cite{valenzuela}).
When more than two levels are involved,  `intrawell'  relaxation  in a time scale $t_i$ is possible.
In the experiments,   $t_i\sim 50{\rm ns}\ll t_r$ \cite{valenzuela}.
This explains that $P_R(t_{\rm exp})\approx {\overline P_R}$, since intrawell processes
dominate relaxation to the asymptotic state and $t_i<t_{\rm exp}$.
(D) For $f_{ac}$ such that the symmetrically located (with respect to $f=1/2$)
avoided crossing between $E_1$ and $E_2$ is reached,
 there are new LZS oscillations.
Furthermore the levels $E_0$ and $E_3$ are also
involved in the dynamics (since the
avoided crossings between  $E_2\leftrightarrow E_3$ and $E_0\leftrightarrow E_1$ 
are traversed by the driving) 
and a more complex dependence of $P_R$ with $f_{ac}$ emerges.
In this case, we  find population inversion,
which is also observed experimentally in \cite{valenzuela}.
A multilevel extension of the semiclassical model of \cite{oliver} can
describe this behavior as well \cite{wen}.

\begin{figure}[h]
\begin{center}
\includegraphics[width=0.9\linewidth,clip]{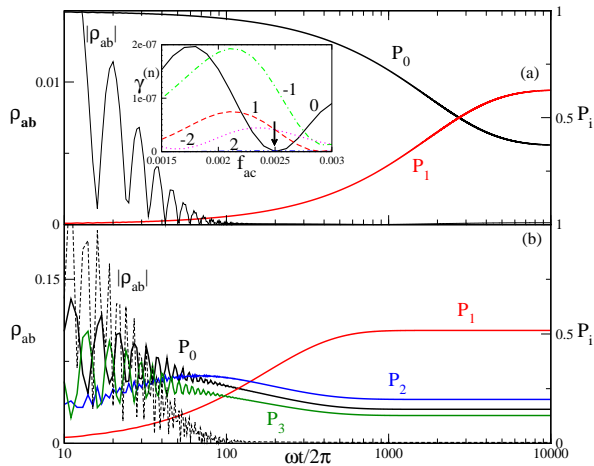}
\end{center}
\caption{(color online)
Off-diagonal (in Floquet basis) density matrix element $\rho_{ab}$ (left axis) 
and level population $P_i$ (right axis) as a function of time (log) for 
$f_{dc}=0.50151$, $\omega=0.003 E_J/\hbar$, $\gamma=0.001$.  
(a) $f_{ac}=0.00245$ and (b) $f_{ac}=0.0191$.
Inset: $n$-photon relaxation terms $\gamma^{(n)}$ 
(the corresponding $n$ value is indicated on the top of each curve) 
in the neighborhood of  $f_{ac}=0.00245$ (indicated by the arrow).} \label{f2}
\end{figure}

In Fig.2 we show   the  explicit time dependence
of $\rho_{\alpha\beta}(t)$ 
for two values of $f_{ac}$ in regimes (B) and (D) respectively.
The off-diagonal elements  $\rho_{\alpha\beta}$ decay
exponentially to zero in a time scale  $t_\phi$ [see Fig.2(a) and (b), left axis].
In the Floquet basis, after full decoherence, 
the relaxation process is determined by the evolution
of the diagonal  $\rho_{\alpha\alpha}(t)$ for $t>t_\phi$ \cite{comment}. 
Thus the asymptotic regime can be obtained 
from the non-trivial solution of 
$\sum_\beta L_{\beta\beta,\alpha\alpha}\,\rho_{\beta\beta}=0$  
after imposing $\frac{d\rho_{\alpha\alpha}}{dt}=0$ in Eq.(2).
In this way, we calculated the   ${\overline P_R}$
shown in Fig.1  as a function of  $f_{ac}$.
In order to understand the emergence of  population inversion
in the  asymptotic limit,
we evaluate  the population of the eigenstates of $H_0$, 
$P_i (t)=\langle \Psi_i|\hat \rho(t)|\Psi_i\rangle$.
We distinguish two mechanisms of PI:

{\it (i) Third-level-mediated population inversion}: 
For large $f_{ac}$ [regime (D) in Fig.1], the  populations $P_i$ relax  to the asymptotic regime 
in a short time scale similar to  $t_i \gtrsim t_\phi$. 
In Fig.2(b) we see that the populations
 $P_2, P_3$ start increasing, and  later they decrease transferring
population to $P_1$. This leads to population inversion, $P_1 > P_0$,  mediated by 
$E_2$ and $E_3$,  which is the usual mechanism for this effect.
In our case the higher levels are reached through  LZS transitions\cite{valenzuela,wen}
instead of resonant transitions.

{\it (ii) Bath-mediated population inversion}:

In Fig.2(a) we show  results for a value of  $f_{ac}=0.00245$ such that only LZS interference
between the two lowest energy levels occurs [regime (B) in Fig.1]. 
Thus one can  restrict to two Floquet states $\alpha=a,b$.
We find that  the decay time $t_\phi$ of the off-diagonal $\rho_{ab}$ is similar
to the previous case but the
level populations
relax to their asymptotic value in a very large time scale $t_r\gg t_\phi$.
Only for $t\gtrsim t_{r}$ population inversion can arise. The large $t_r$
explains the difference between ${\overline P_R}$
and the finite time $P_R(t\sim t_{\rm exp})$ in Fig.1.
To understand the  underlying mechanism  we decompose the relaxation 
rate $\gamma_r={t_{r}}^{-1}$  as  $\gamma_r = 2(R_{aabb}+R_{bbaa})= \gamma^{(0)} +\sum_{n\not=0} \gamma^{(n)}$,
with $\gamma^{(n)}=2(g_{ab}^n|\Gamma_{ab}^n|^2+g_{ba}^n|\Gamma_{ba}^n|^2)$ the terms that
describe virtual  n-photon transitions to bath oscillator
states  \cite{grifonih}. In the inset of  Fig.2(a) we show that 
when there is PI the term $\gamma^{(n=-1)}$  is   the largest  while 
$ \gamma^{(n=0)}\approx 0$.
This indicates that the dominant mechanisms leading  to  PI
is a transition to a virtual level at energy  $E_0+\hbar\omega > E_{1}$ (one  photon absorption, $n=-1$),
followed by a relaxation to $E_{1}$.
Previous works have found PI,
under various approximations, 
but for the asymptotic regime of TLS \cite{popin2l,hartmann,stace2005,yu2010a}.
However the time-dependent dynamics with different time scales 
has been overlooked.  In fact, the relevant point from our
findings is that the asymptotic regime  is difficult
to reach in the experiment, since PI  needs
long times ($t\gtrsim t_r\gg t_\phi$) to emerge when mediated by the bath.
Moreover, the difference between $P_{R} (t_{\rm exp})$  and  the asymptotic $\overline P_R$, 
is enhanced when decreasing temperature (see inset of Fig.1) 

\begin{figure}[h]
\begin{center}
\includegraphics[width=0.9\linewidth,clip]{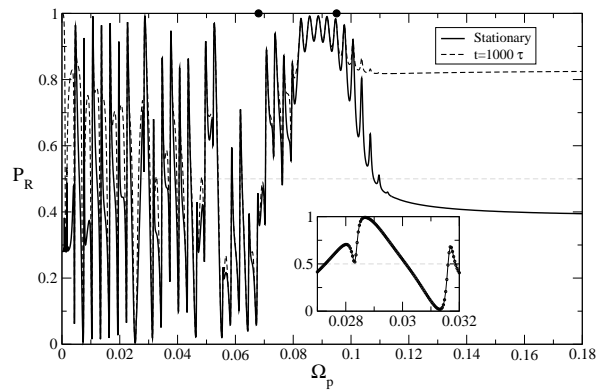}
\end{center}
\caption{
Population $P_R$ as a function of the  frequency 
$\Omega_p$ (in units of $E_J/\hbar$) for 
$T=20{\rm mK}$, $f_{dc}=0.50151$, 
$f_{ac}=0.00245$, $\omega=0.003E_J/\hbar$, $\gamma=0.001$ and $\kappa=0.001$. 
Continuous line shows the asymptotic $\overline P_R$  and dashed line corresponds 
to $P_R(t= 1000 \tau \sim t_{\rm exp})$. 
The circles indicate the values of $\Delta_+$ and $\Delta_-$. The inset shows 
$\overline P_R$ as a function of $\Omega_p$ around the $n=9$ resonance. 
See text for details.} \label{f3}
\end{figure}

An interesting consequence of the `bath-mediated' mechanism
is that it enables - by changing the spectral structure of the bath -
to  modify the asymptotic state of the FQ.
In particular, a realistic modeling of the
environmental bath for FQ  includes 
a read-out dc SQUID inductively
coupled to the FQ \cite{qbit_mooij,caspar,chiorescu2}. 
For this case, the   bath  spectral density  becomes
$J_{sb}(\omega)={\gamma \omega\Omega_p^4}/
[(\Omega_p^2-\omega^2)^2+(2\pi\kappa
\omega\Omega_p)^2]$, with $\Omega_p$ the resonant frequency
of the SQUID detector,
and $2\pi\kappa\Omega_p$ the width of the resonance at 
$\omega= \Omega_p$ \cite{caspar}.
In Fig.3 we show  ${\overline P_R}$
 as a function of $\Omega_p$ for the same
$f_{ac}$ of Fig.2(a), but after solving Eq.(2)  with $J_{sb}(\omega)$. 
We identify three regimes in the 
overall behavior of ${\overline P_R}$ vs $\Omega_p$ that we describe below.
Defining $\Delta(f)= E_1(f)-E_0(f)$, $\Delta_+=\Delta(f_{dc}+f_{ac})$
and $\Delta_-=\Delta(f_{dc}-f_{ac})$ we get:
(1) {\it Multiphoton resonances with the bath}.
 For $\Omega_p < \Delta_{-}$ the population ${\overline P_R}$ has a strong
dependence on $\Omega_p$, displaying resonances for  $\Omega_p \approx n \omega
+ \varepsilon_b-\varepsilon_a$ (with $\varepsilon_b, \varepsilon_a$ 
the quasienergies of the Floquet states mostly weighted 
in  the two lowest $H_0$ eigenstates).
These resonances correspond to the maxima at $\Omega_p$ of the
$J_{sb}(\varepsilon_a-\varepsilon_b+n\hbar \omega)$, 
and have been previously discussed for $n=1$  
within a perturbative approach \cite{milenas}.
Near these resonances it
is possible to fully control the population ${\overline P_R}$, 
 going from ${\overline P_R}\sim 1$ to complete
population inversion ${\overline P_R}\sim 0$,
with small changes of $\Omega_p$. In the inset of Fig.3 we show 
in detail the case near the $n=9$ resonance.
(2) {\it Bath mediated population inversion}.
 In the opposite limit, $\Omega_p > \Delta_+$, the behavior
is similar to the one discussed previously in Fig.2(a), 
since for large $\Omega_p$, the spectral density $J_{sb}(\omega)\rightarrow \gamma \omega$
(3) {\it Bath mediated cooling}.
In the intermediate regime, $\Delta_{-} \lesssim \Omega_p \lesssim \Delta_+$, 
transitions to an
effective  energy level \cite{milenas,chiorescu2} at $E_0 + \Omega_p < E_1$
predominate. From this effective level it
is possible to decay and repopulate the ground state, 
obtaining  ${\overline P_R}\sim 1$.
The resonances at $\Omega_p \approx n \omega
+ \varepsilon_b-\varepsilon_a$ are still observed. 
A full picture of the effect of a structured bath can be observed in Fig.4,
which shows a  map of the asymptotic population ${\overline P_R}$ as a function of $\Omega_p$ and
$f_{ac}$.
The onset of third level mediated PI at higher $f_{ac}$ 
(corresponding to the second 'diamond' of \cite{valenzuela}) is also observed.
Notice that this type of PI is  almost independent of $\Omega_p$,
while the bath-mediated mechanism is active for $\Omega_p > \Delta_+$.

\begin{figure}[h]
\begin{center}
\includegraphics[width=0.9\linewidth,clip]{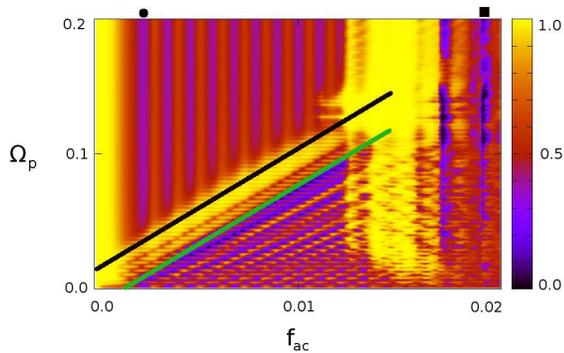}
\end{center}
\caption{(color online)
 Contour map of the asymptotic population 
$\overline P_R$ as a function of the driving amplitude $f_{ac}$ and the resonance frequency $\Omega_p$ 
(in units of $E_J/\hbar$) for 
$T=20{\rm mK}$, $f_{dc}=0.50151$, $\omega=0.003E_J/\hbar$, $\gamma=0.001$ and $\kappa=0.001$. 
The black and green lines represent the positions of $\Delta_+$ and $\Delta_-$.
Black circle: $f_{ac}$  corresponding to Fig.2(a) and Fig.3;
black square: $f_{ac}$ corresponding to Fig.2(b).
} \label{f4}
\end{figure}

In conclusion, by performing a realistic modeling of the 
FQ we are able to analyse different scenarios
for population inversion in strongly driven quantum systems,
understanding the parameter regimes for their occurrence.
One puzzling situation is that the bath-mediated PI discussed here
has not been observed in  \cite{valenzuela}.
The large time scale separation $t_r \gg t_\phi$ 
in the device of \cite{valenzuela} suggests that
the  PI could be beyond the experimental time window \cite{comment3}.
Indeed, our results explain that to observe this effect 
the experiments should be carried out
for long driving times,
after full relaxation with the bath degrees of freedom ($t\gtrsim t_r$).
A remarkable point we found is the dependence of the
LZS oscillations on the frequency $\Omega_p$ of the
SQUID detector. The FQ fabricated with 
Nb junctions as in \cite{oliver,valenzuela} have typically
a small gap and thus they are in the regime $\Omega_p > \Delta_+$,
where bath-mediated PI can take place.
The FQ fabricated with Al junctions, as in \cite{chiorescu,chiorescu2},
have a large gap and thus they are expected to be in the regime
$\Omega_p < \Delta_-$. Amplitude
spectroscopy measurements  carried out in
these later systems could give different `diamond' patterns as a function
of $f_{dc},f_{ac}$ with resonances and LZS interference effects
from the oscillator levels at $E_i+n\hbar\Omega_p$. 
In principle, the frequency $\Omega_p$ can
be controlled  by varying in the SQUID detector the driving
current or the shunt capacitor \cite{chiorescu2},
or by adding a tunable resonator to the circuit \cite{deppe}.
In these cases there is room to  fully tailor the
population $P_R$. Furthermore, 
the discussed population inversion scenarios
could also apply to
other quantum systems in which LZS interference
has been measured \cite{cqubits,yoakum,mark,kling,huang}.

We acknowledge discussions with Sergio Valenzuela
and financial support from CNEA,  CONICET
(PIP11220080101821 and PIP11220090100051) and ANPCyT (PICT2006-483
and PICT2007-824).


\begin{thebibliography}{}
\bibitem{comment3} Moreover, $t_r/t_\phi$ in \cite{valenzuela}
is much larger than for the parameters used here, where for numerical convenience
we have chosen  a  larger gap $\Delta$ and a larger $\gamma$.
\end{thebibliography}
\end{document}